\documentclass[a4paper,11pt]{article}
\usepackage{a4wide}
\usepackage{graphics}
\usepackage{epsf}
\newcommand{\ve}{\ensuremath{\varepsilon}}

\title{Kinetic Monte Carlo Simulations of dislocations in heteroepitaxial growth}
\author{F. Much\footnote{e-mail: {\tt much@physik.uni-wuerzburg.de}}, M. Ahr, M. Biehl and W. Kinzel \\
Sonderforschungsbereich 410 \\
Institut f\"{u}r Theoretische Physik \\
Julius-Maximilians-Universit\"{a}t 
W\"{u}rzburg\\
 Am Hubland, 97074 W\"{u}rzburg, Germany }

\begin{document}
\maketitle
\setlength{\unitlength}{\textwidth}

\begin{abstract}

We determine 
the {\it critical layer thickness} for the 
appearance of misfit dislocations as a function of the misfit
$\varepsilon$  between the lattice constants of the substrate and the adsorbate
from Kinetic Monte Carlo (KMC) simulations of heteroepitaxial growth. \\
To this end, an algorithm is
introduced which allows the off-lattice simulation of 
various phenomena observed in heteroepitaxial growth 
(see {\it e.g.} \cite{politi00, pimpinelli}) including {\it critical layer 
thickness} for the appearance of misfit dislocations, or 
{\it self-assembled} island formation.  
The only parameters of the model are deposition flux,
temperature and a pairwise interaction potential between
the particles of the system. \\
Our results are compared
with a theoretical treatment of the problem and show good agreement with a 
simple power law.
\end{abstract}

\section{Introduction}
Heteroepitaxial growth has been a field of intense study in recent years. 
This is mainly due to the improved performance of semiconductor and 
opto-electronic devices which can be achieved using strained layer epitaxy. 
Here the heteroepitaxial growth by depositing material onto a
substrate with the same crystal structure as the adsorbate but a slightly different lattice 
constant is of special interest.   
\par
In the early stages of this kind of heteroepitaxial growth the adsorbate is
coherent with the substrate. In this state the crystal topology is
that of a perfect crystal, {\it i.e.} each particle has the same coordination
number and its nearest and next-nearest neighbors form the same 
geometrical figure with only slightly modified distances \cite{politi00, pimpinelli}.
\par
As the thickness of the adsorbate film increases the elastic
energy of the film rises until it is energetically favorable to form 
dislocations in order to relieve the strain. 
In this new incoherent state the crystal topology is perturbed near the 
substrate/adsorbate interface. The thickness of the adsorbate film at which this
occurs is called the {\it critical layer thickness} $h_{c}$.\\
Theoretical models were proposed in order to determine $h_{c}$ 
as a function of material parameters \cite{politi00, matthews, Dong97, cohen94}.
In this letter we determine the critical layer thickness as a function of the misfit 
between the lattice constants of the substrate and the adsorbate using computer simulations.
\par
In order to simulate heteroepitaxial growth one has to overcome the limitations of a fixed lattice. 
One obvious way to perform off-lattice simulations are molecular dynamics (MD), 
see for instance \cite{Dong97} in the context of misfit dislocations. MD methods have the advantage of conceptual simplicity but 
can only be applied to rather small system sizes at very high temperature and deposition rates. 
Because of the high computational efforts 
the critical layer thickness has been determined for only a few values 
of the misfit.
\par
Here, we propose a KMC algorithm for the simulation of the early stages of heteroepitaxial growth. 
In contrast to similar off-lattice algorithms suggested before 
- for example by Faux {\it et al.} \cite{Faux90, Faux94} , Plotz {\it et al.} \cite{Plotz92} 
 or Schindler  \cite{schindler00} - we are able to simulate heteroepitaxial growth for rather 
thick adsorbate layers and over a wide range of the misfit between the lattice constants 
of the substrate and the adsorbate.  
\par
In the following we consider only growth in $1+1$ dimensions. However, the method can be extended to 
$2+1$ dimensional growth. 
To our knowledge this is the
first time that the critical layer thickness for the appearance of dislocations is observed in 
Monte Carlo simulations. We find that our results fit well to a power law, as it has been observed for 
several semiconductor compounds \cite{cohen94}.

\section{Method}
The aim here is to gain 
general insight into relevant mechanisms of heteroepitaxial growth. We 
therefore choose a simple {\it Lennard-Jones} potential
\begin{equation}
\label{interaction_potential}
U_{ij}(\sigma)=4U_{0}\left[\left(\frac{\sigma}{r_{ij}}\right)^{12}-
\left(\frac{\sigma}{r_{ij}}\right)^{6}\right].
\end{equation}
as particle interaction, which is numerically easy to handle and saves computer time 
compared to more realistic empirical potentials like the EAM approximation 
(see {\it e.g.} \cite{Somfai99}). However, we focus on the observation of effects which should not 
depend on the particular choice of the potential. 
\par 
The equilibrium distance $r_{0}$ between two isolated particles interacting 
via $U_{ij}$ becomes $r_{0}=\sqrt[6]{2}\sigma$ and is slightly smaller in 
the bulk material. Because of the isotropy of the Lennard-Jones potential
the particles arrange in a triangular lattice. 
In order to save computer time the interaction potential 
$U_{ij}$ is cut off at a distance $r_{ij}>3r_{0}$. 
The interaction strength at this distance is 
less than $1\%$ of the value at the equilibrium distance and can therefore be neglected. 
The interaction of two substrate particles is given by $U_{ij}\left(\sigma_{S}\right)$. Two adsorbate 
particles interact via $U_{ij}\left(\sigma_{A}\right)$ whereas we assume that  
a substrate and an adsorbate particle interact via  
$\frac{1}{2}\left(U_{ij}\left(\sigma_{S}\right)+U_{ij}\left(\sigma_{A}\right)\right)$.
\par
We would like to stress that growth is not simulated on a fixed lattice 
but rather two particles $i$ and $j$ are separated by a {\it continuous} distance $r_{ij}$. 
There are two possible events in our simulations: deposition and diffusion of {\it adsorbate} 
particles. The two-dimensional simulation cell is open in
vertical and has periodic boundary conditions in lateral direction.
Adsorbate particles are randomly deposited on the crystal surface with a rate
$R_{d}=LF$,where $L$ is the system size and $F=1s^{-1}$ is the deposition flux.
The rate $R_{i}$ for a diffusion event $i$ is given
by an Arrhenius law 
\begin{equation}
\label{diffusion_rate}
 R_{i}=\nu_{0}e^{-\frac{E_{a,i}}{k_{B}T}},
\end{equation}
where $\nu_{0}=10^{12}s^{-1}$, $E_{a,i}$, $T$ are the attempt frequency, 
the activation barrier for the diffusion step $i$ and the simulation temperature, 
respectively.\\
$E_{a,i}$ is given by 
$E_{a,i}=E_{t,i}-E_{b,i}$, where $E_{t,i}$ denotes the energy of the particle at the
transition state and  $E_{b,i}$ the energy at the binding state.
Both are calculated for a frozen crystal using Brent's method \cite{num_rec92}.
Here the saddle point search for the calculation of $E_{t,i}$ can be replaced 
by a simpler maximum search in $1+1$ dimensions. \\
To consider the elastic deformation of the crystal
after each microscopic event (diffusion or deposition)
the total potential energy 
of the $n$ particles - system 
\begin{equation}
\label{total_energy}
E_{tot}=\sum_{i=1}^{n}\sum_{j=i+1}^{n}U_{ij}
\end{equation}
is minimized using a conjugate gradient method \cite{num_rec92}
under variation of the coordinates of all particles (substrate and adsorbate) 
within a circle of radius $3r_{0}$ around the particle
where the event took place. 
In order to  avoid strain caused by this local relaxation
of the crystal, after a distinct number of microscopic
events - depending on the misfit $\varepsilon$ -  
a minimization of $E_{tot}$ under variation of all particle coordinates  
is performed.
\par
The obtained rates for deposition and diffusion of adsorbate particles are used in a 
rejection free KMC simulation.
Using a binary tree structure \cite{new99} an event $i$ is chosen
with the correct probability $R_i/R$, where
\begin{equation}
\label{total_rate}
 R=R_{d}+\sum_{i}{R_{i}}
\end{equation}
is the total rate of all microscopic processes. Then this event is performed and the rates
of all affected events are updated. Unlike in standard Monte Carlo simulations
time does not advance linearly in discrete time steps $\Delta t$. Instead the time 
interval $\tau$ between two microscopic processes is randomly drawn from a Poisson distribution
$P(\tau)=Re^{-R\tau}$ by $\tau=-\frac{\ln\rho}{R}$,
where $\rho$ is a uniformly distributed random number between 0 and 1. 
\par
Each simulation run starts with six atomic layers of substrate with a fixed 
bottom layer. The system size $L$ (number of particles in the substrate's upper 
layer) is between $L=100$ and $L=200$. 
Within this range we found no significant dependence of the results on $L$. 
Measuring lengths in units of $\sigma{_S}$, 
$\sigma_{A}$ is chosen between $0.85$ and $1.11$, so we can simulate 
heteroepitaxial growth for misfits
\begin{equation}
\label{misfit}
\varepsilon=\frac{\sigma_{A}-\sigma_{S}}{\sigma_{S}}
\end{equation}
between $-15\%$ and $+11\%$. 
 
\section{Results and Discussion}

\begin{figure}
\centering

 \begin{minipage}[t]{0.48 \textwidth} 
 \epsfxsize= 0.79 \textwidth
 \epsffile{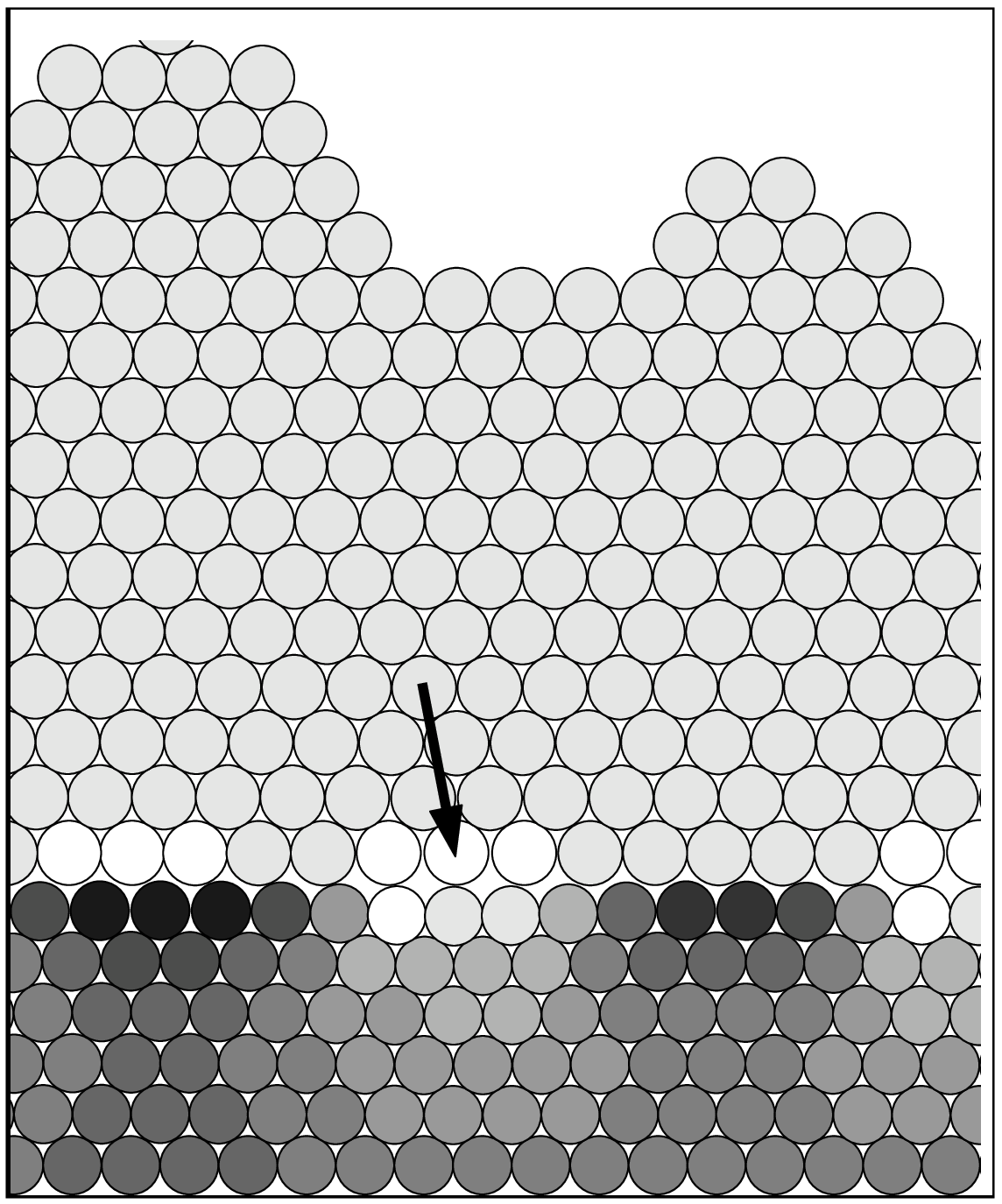}
 
 \end{minipage}
  \centering
 \begin{minipage}[t]{0.48 \textwidth} 
 \epsfxsize= 0.79 \textwidth
 \epsffile{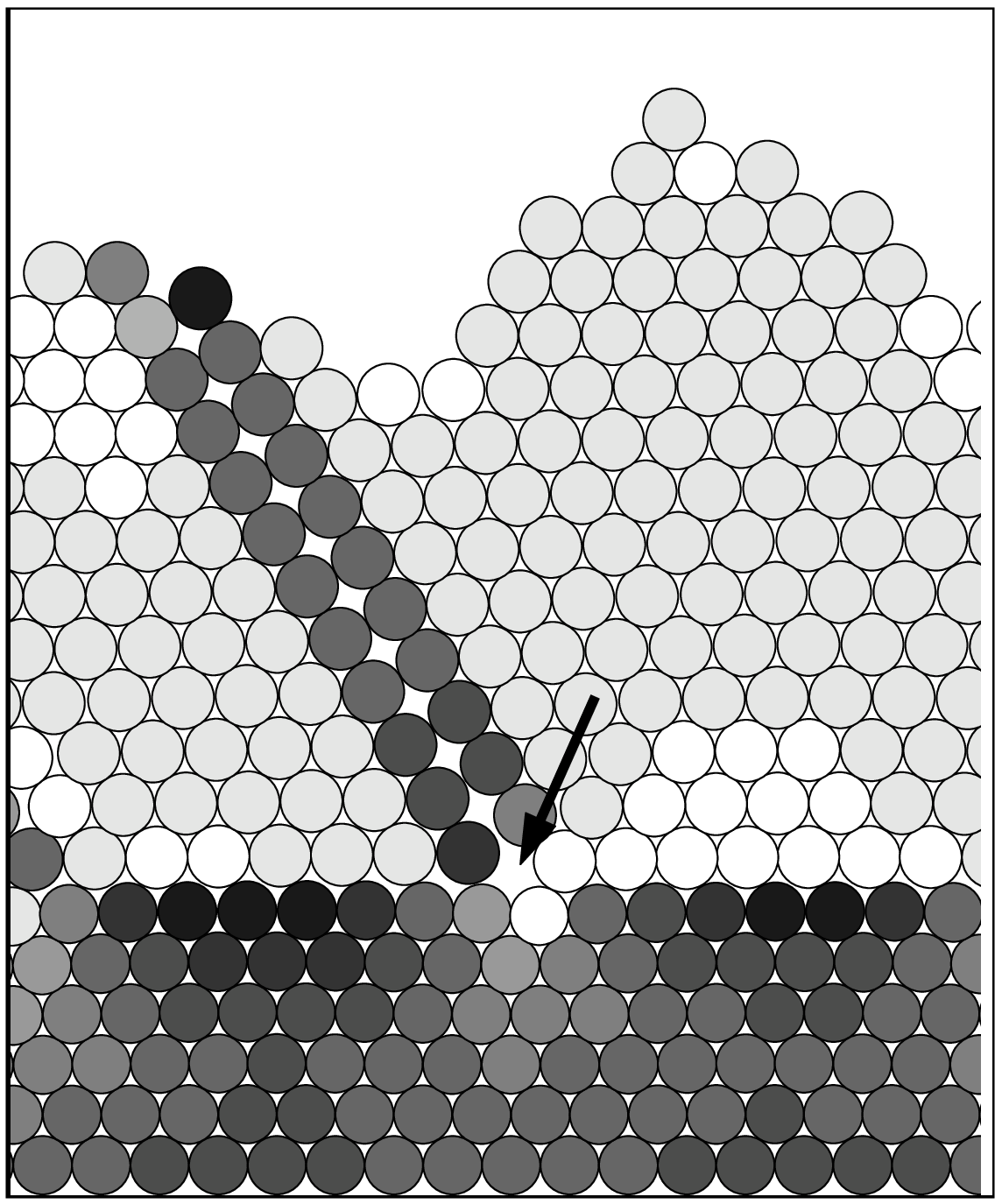} 
 
\end{minipage}
 \caption{Typical sections of crystals obtained in our simulations. The six bottom layers
          are the given substrate. The dislocations are marked with arrows. 
          Left panel: perfect dislocation for $\varepsilon=+10\%$. Right panel: partial dislocation for
          $\varepsilon=+6\%$. 
          The grey level for a particle indicates the particle's average distance to its nearest 
          neighbors of the same kind: the lighter its grey level the more is this particle under 
          compression. }
 \label{pandp}

\end{figure}

In the following the temperature is set to $T=0.03 \frac{U_{0}}{k_{B}}$, which is
sufficiently low compared to the melting temperature $T_{m}=0.415 \frac{U_{0}}{k_{B}}$
of a pure two-dimensional Lennard-Jones solid \cite{abraham}.
\par
Heteroepitaxial growth is now simulated in order to determine
the critical layer thickness $h_{c}$ for the appearance of 
dislocations as a function of the misfit \ve.\\
To this end between 5 and 10 independent simulation runs are carried out for each value of $\varepsilon$.
We find that in each simulation run several dislocations appear simultaneously. 
Then, after the deposition of a few monolayers of adsorbate after the first 
appearance of a dislocations in the crystal the number of dislocations remains constant. 
The thickness of the adsorbate layer at which dislocations first appear is registered as $h_c$.
\par
To prove the existence of a dislocation we determine the coordination number $n_{c}$ of each particle 
by calculating the Voronoy polyhedra \cite{ash76, Strandburg88}. 
Voronoy polyhedra are a generalization of the 
Wigner-Seitz cell to a system without a fixed lattice. The number of sides of a Voronoy polyhedron gives the 
coordination number $n_{c}$ ($n_{c}=6$ for a particle in a perfect triangular lattice). 
A Burgers circuit \cite{tod68} is drawn  
around regions of the crystal with $n_{c} \not= 6$. 
A non-vanishing Burgers vector then indicates the appearance of a dislocation.
\par
Figure \ref{pandp} shows sections of two crystals obtained in our simulations for 
left panel $\varepsilon=+10\%$ and right panel $\varepsilon=+6\%$. 
The grey level for a particle in these pictures is obtained from the 
particle's average distance to its nearest neighbors of the same kind. 
The lighter the grey level the more is this particle under compression.

\subsection{Number of Dislocations}
\begin{figure}
\centering
 \begin{minipage}[t]{0.45 \textwidth} 
 \epsfxsize= 0.98 \textwidth
 \epsffile{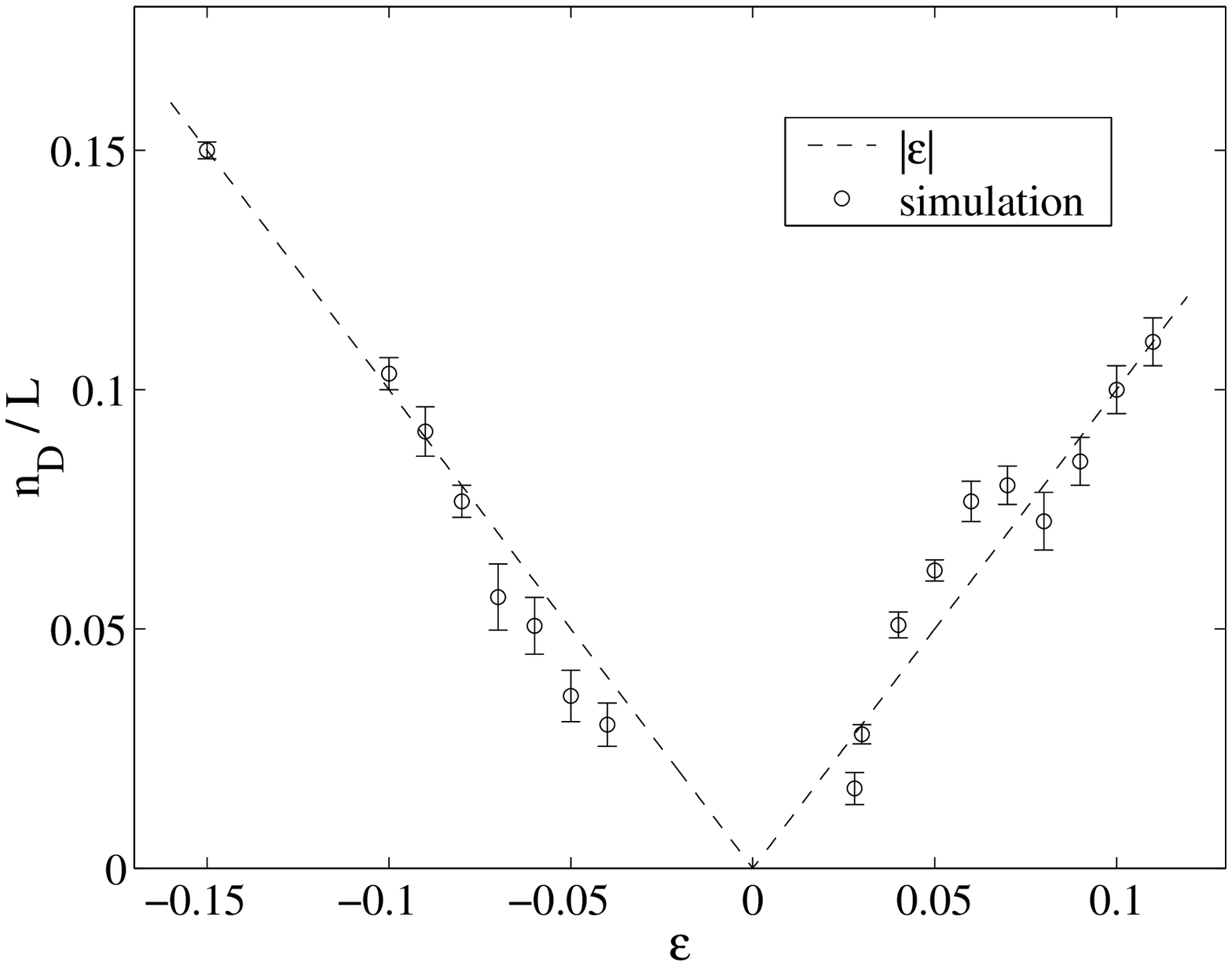}
 \caption{Number of dislocations per system size $n_{D}/L$ as a function of the misfit \ve. 
          The error bars represent as the standard error  
          of the simulation results.
          The dashed line gives the theoretical number of {\it perfect} 
          dislocations in a system of size $L$.}
 \label{diszahl}
 
 \end{minipage}
 \hspace{2em}
 \begin{minipage}[t]{0.45 \textwidth} 
 \epsfxsize= 0.99 \textwidth
 \epsffile{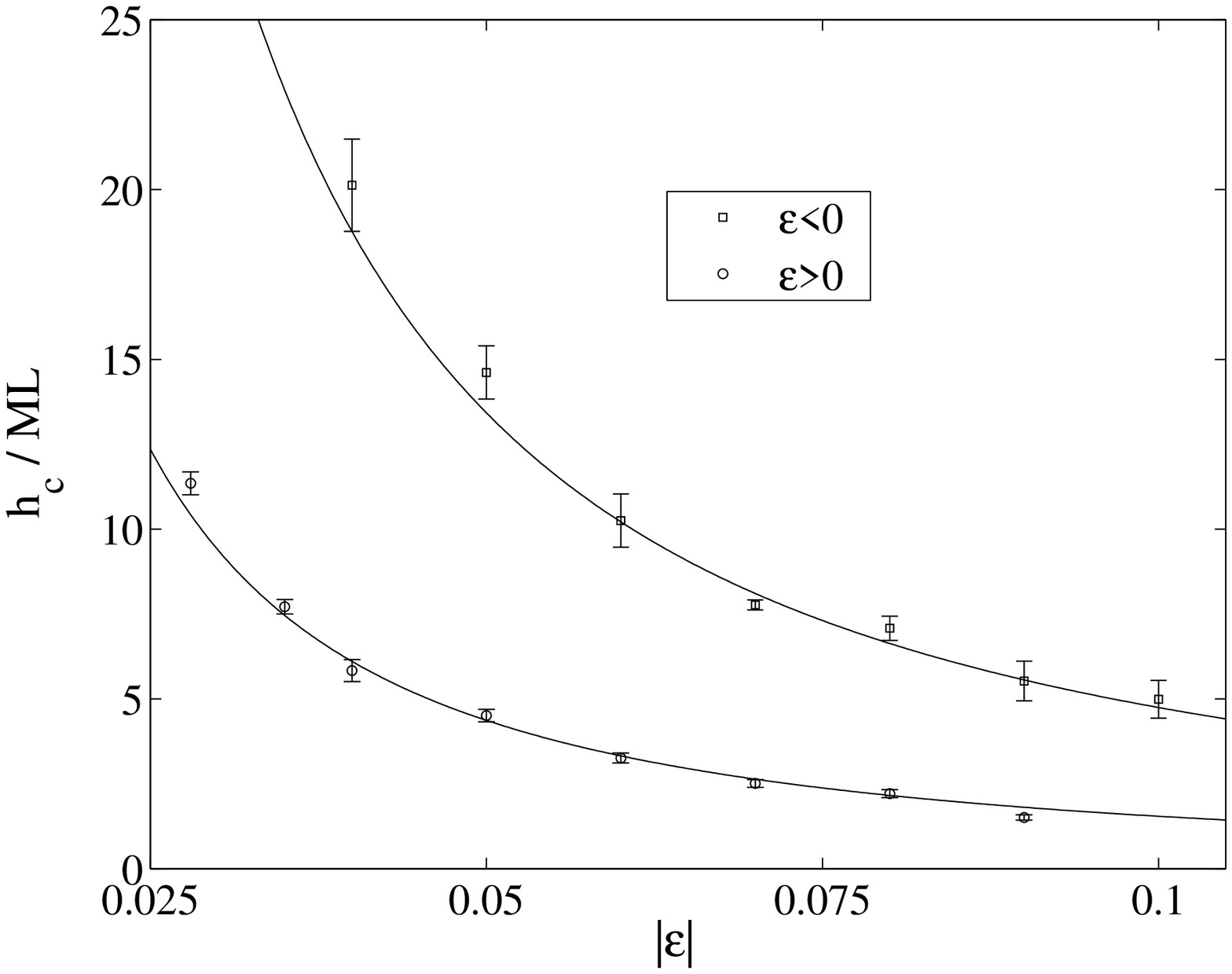}
 \caption{Critical thickness $h_{c}$ versus misfit $|\ve|$ for
  $\ve<0$ (upper curve) and $\ve>0$ (lower curve). The error bars are obtained as the standard error  
  of the simulation results. The solid lines are 
  calculated using eq. (\ref{theory}) where $a^{*}=0.15$ for $\ve<0$ and 
  $a^{*}=0.05$ for $\ve>0$. }
 \label{loglogn}
\end{minipage}
\end{figure}

Figure \ref{diszahl} shows the number of dislocations 
per unit length $n_{D}/L$ counted for each value of $\varepsilon$  
about 6 monolayers deposited adsorbate after the first 
appearance of a dislocation in the crystal (the maximum number of dislocations 
should be reached for this thickness of the adsorbate layer). 
 The dashed line gives the theoretical 
number of dislocations in a system of size $L$  under the assumption 
that $n_{D}=L|\ve|$ perfect dislocations can appear. {\it Perfect dislocations} (fig. \ref{pandp} left panel)  
are those for which the crystal topology far from the substrate/adsorbate interface is the same
as in the coherent state and the Burgers vector is therefore an integer multiple of the lattice
vector.
The formation of partial dislocations (fig. \ref{pandp} right panel) 
- characterized by a Burgers vector which is a   
rational fraction of a lattice vector - causes the 
deviations from the theoretical results for $-0.07 \leq \ve \leq -0.03$ 
and $0.04 \leq \ve \leq 0.08$. This is due to the fact, that partial dislocations are spatially more 
extended than perfect dislocations. For this reason in the case $\varepsilon>0$ more and for $\varepsilon<0$ 
less dislocations than $n_{D}=L|\ve|$  have to be built when partial dislocations appear. 
Why partial dislocations 
only appear for distinct values of $\varepsilon$ is still unknown.

\subsection{Critical layer thickness}

Figure \ref{loglogn} shows the critical layer thickness $h_{c}$ 
plotted versus the absolute value of the misfit \ve. 
For $-0.03 < \ve < 0.02$ the critical thickness is too large 
to be observed in our simulations. \\ 
The simulation results show a dependence of $h_{c}$ on the sign of 
the misfit. This was found before by L. Dong {\it et al.} \cite{Dong97}. We believe this dependence
is due to the fact, that the Lennard-Jones potential is not harmonic. 
The potential is steeper in compression ($\ve>0$) than in tension ($\ve<0$), so that for $\ve>0$
it becomes favorable to form a dislocation for smaller values of $h_{c}$. 
\par
Our simulation results agree well with a power law (solid lines in fig. \ref{loglogn}) 
\begin{equation}
\label{theory}
 h_{c}=a^{*} \ve^{-3/2}
\end{equation}
which was proposed by Cohen-Solal {\it et al.} \cite{cohen94}. \par
There, an energy balance 
model is proposed for calculating the critical layer thickness in heteroepitaxial growth of semiconductor 
compounds. To this end the classical strain energy, without any change of the substrate 
or dislocation formation, 
and the deformation energy due to a full system of interfacial misfit dislocations were compared. 
The method yields the $\ve^{-3/2}$ power law, where $h_c$ depends mainly on the misfit $\varepsilon$. 
Their model was compared with experimental
data revealing an excellent agreement for IV-IV, III-V and II-VI semiconductor compounds
with values for $a^{*}$ between $a^{*}=0.15$ and $a^{*}=0.50$. A nonlinear fit of our results yields 
$a^{*}=0.15$ for $\ve<0$ and $a^{*}=0.05$ for $\ve>0$

\section{Conclusion}

The goal of this examination was to determine the critical layer thickness 
for heteroepitaxial growth of Lennard-Jones particles. For this purpose we propose a novel 
KMC algorithm. \\
We demonstrate, that with this algorithm it is possible to simulate the 
appearance of dislocations for a wide range of misfits and rather large system sizes 
compared to commonly used methods like molecular dynamics simulations. \par
We find our simulation data in good agreement with a very simple 
$\varepsilon^{-3/2}$ power law, where the critical thickness depends mainly 
on the lattice mismatch. This law has been identified in various systems before \cite{cohen94}
 and may thus be considered as quite general. 
\par

The developed algorithm is applicable in the simulation of various other phenomena observed in 
heteroepitaxial growth.
In future work we will examine the 2D-3D transition in island growth and step-bunching on 
vicinal surfaces, for instance. \\

We would like to thank A. Schindler for fruitful discussions about problems concerning the 
simulation of heteroepitaxial growth. FM and MA were supported by the Deutsche Forschungsgemeinschaft.


\begin{thebibliography}{0}

\bibitem{politi00} 
  Politi P., Grenet G., Marty A., Ponchet A. and Villain J., {\it Phys. Rep.} {\bf 324} 271, 2000

\bibitem{pimpinelli} 
  Pimpinelli A. and Villain J., {\it Physics of Crystal Growth}, Cambridge University Press, 1998

\bibitem{matthews} 
  Matthews J. W. and Blakeslee A. E., {\it J. Crystal Growth} {\bf 27} 118, 1974

\bibitem{Dong97} 
 Dong L., Schnitker J., Smith R. W. and Sroloviz D. J., {\it J. Appl. Phys.} {\bf 83} 217, 1997

\bibitem{cohen94} 
 Cohen-Solal G., Bailly F. and  Barb\'{e} M., {\it J. Crystal Growth} {\bf 138} 68, 1994

\bibitem{Faux90} 
 Faux D. A., Gaynor G., Carson C. L., Hall C. K. and Bernholc J., {\it Phys. Rev. B} {\bf 42} 2914, 1990

\bibitem{Faux94}
 Spjut H. and Faux D. A., {\it Surf. Sc.} {\bf 306} 233, 1994

\bibitem{Plotz92} 
 Plotz W. M., Hingerl K. and  Sitter H., {\it Phys. Rev. B} {\bf 45} 12122, 1992

\bibitem{schindler00} 
 Schindler A., {\it Theoretical aspects of growth on one and two dimensional strained crystal surfaces}, Dissertation, Duisburg 1999

\bibitem{Somfai99} 
  Somfai E. and Sander L. M., {\it Strain in heteroepitaxial growth}, Cond-mat/9909328, 1999

\bibitem{num_rec92}
 Press W.H., Teukolsky S. A., Vetterling W. T. and  Flannery B. P., {\it Numerical Recipes in C}, Cambridge University Press, 1992

\bibitem{new99} 
 Newman M. E. J. and Barkema G. T., {\it Monte Carlo Methods in Statistical Physics}, Oxford University Press, 1999


\bibitem{abraham} F.F. Abraham, in {\it Melting, Localization, and Chaos}, 
edited by R.K. Kalia and P. Vashishta. Elsevier, New York, 1982. 

\bibitem{ash76} N.W. Ashcroft, N.D. Mermin. {\it Solid State Physics}. Saunders College Publishing, 1976.

\bibitem{Strandburg88} K. J. Strandburg. {\it Rev. Modern Physics} {\bf 60}(1) 1988

\bibitem{tod68} J. P. Hirth, J. Lothe, {\it Theory of Dislocations}. McGraw-Hill Book
Company, 1968.






\end{thebibliography}
\end{document}